\documentclass[floatfix,aps,prl,showpacs,amsmath,nofootinbib,
preprintnumbers,twocolumn]{revtex4}
\usepackage{graphicx}

\def\be{\begin{equation}}
\def\ee{\end{equation}}
\def\bea{\begin{eqnarray}}
\def\eea{\end{eqnarray}}

\begin{document}

%\preprint{astro-ph/yymmddd}

\title{Inflation with violation of the null energy condition}

\author{Marco Baldi}
\email{baldi@bo.iasf.cnr.it}
\affiliation{IASF-Bologna/INAF, 
Istituto di Astrofisica Spaziale e Fisica Cosmica, Istituto Nazionale di 
Astrofisica, 
via Gobetti, 101 -- I-40129 Bologna -- Italy}

\author{Fabio Finelli}
\email{finelli@bo.iasf.cnr.it}
\affiliation{IASF-Bologna/INAF, 
Istituto di Astrofisica Spaziale e
Fisica Cosmica, Istituto Nazionale di Astrofisica,
via Gobetti, 101 -- I-40129 Bologna -- Italy}
%\altaffiliation{Also supported by INFN, Sezione di Bologna.}

\author{Sabino Matarrese}
\email{sabino.matarrese@pd.infn.it}
\affiliation{Dipartimento di Fisica ``G. Galilei'', Universit\'a di Padova,\\ 
and INFN Sezione di Padova, via Marzolo 8, I-35131 Padova, Italy}

\date{\today} 

\begin{abstract} 
Inflation may have been driven by a component which 
violated the
Null-Energy Condition, thereby leading to {\it super inflation}.
We provide the formalism to study cosmological perturbations when  
such a component is described by a scalar field with arbitrary Lagrangian. 
Since the
background curvature grows with time, gravitational waves always have a
blue spectrum. Scalar perturbations may also have a blue spectrum, 
{\em albeit} in single field models. We apply our formalism to the
case of phantom inflation with an exponential potential 
(whose pole-like inflationary stage is an attractor for
inhomogeneous cosmological models for any value of the potential slope) 
as an example. We finally compare the predictions of super inflation 
with those of standard inflation stressing the role
of gravitational waves. 
\end{abstract}


\maketitle
\emph{Introduction}. 
Inflation is the most promising theory of structure formation. 
Its predictions for the simplest case of a single scalar field model 
with a nearly scale-invariant spectrum of 
Gaussian curvature perturbations are in good agreement with observational 
data. 
The ``smoking gun'' for inflation is the detection of the stochastic 
background of relic gravitational waves, 
which carries information on the energy scale of inflation. 
For a scalar field described by a canonical Lagrangian, 
the amplitude $P_T$ and spectral index $n_T$ (at a certain scale $k_0$) 
of gravitational waves are locked to the amplitude of scalar perturbations 
$P_S$ by the {\it consistency relation} $r \equiv P_T/P_S= - 8 n_T$. 
While this relation is violated in multi-field inflationary models, 
Garriga and Mukhanov \cite{GM} showed how it can be violated also in single 
scalar field models with a non-canonical Lagrangian. 

What remains a generic prediction of inflation - single or multi-field 
models - in Einstein theories is a {\it red spectrum} for 
gravitational waves. 
Such a prediction is related to the decrease of the Hubble 
parameter $H$ during inflation \cite{LM2}. 
In this paper we focus on the possibility 
that $H$ may grow during inflation, as it occurs when the Null Energy 
Condition (NEC), {\it i.e.} $p+\rho\geq 0$ (with $p$ the pressure and 
$\rho$ the energy density), is violated. Although known forms of matter 
seem to respect NEC, there is no evidence that the universe as a whole 
does at the present time \cite{knop}: it is then conceivable and  
interesting to explore the consequences of such a violation in the early past, 
i. e. during inflation. For this purpose, we provide the treatment of scalar 
perturbations for general scalar field 
theory, extending previous works \cite{GM}. 
We show that, in contrast to common belief, NEC violating theories can be
completely stable at the classical level when gravitational 
perturbations are self-consistently taken into account. 

The distinctive signature of {\em super inflation} is a {\it blue-tilted} 
spectrum for relic gravitational waves, which offers better chances to 
be detected indirectly in Cosmic Microwave Background (CMB) 
anisotropy measurements. Specific realizations of this 
model may even lead to the possibility of direct detection by 
space-borne interferometers, such as LISA \cite{lisa} or BBO \cite{bbo}. \\

\emph{Basic Equations}. Let us consider the action for gravity plus 
a scalar field $\phi$ with the generic Lagrangian $p(\phi,\chi)$, 
\be
S \equiv \int d^4x {\cal L} = \int d^4x \sqrt{-g} \left[
\frac{R}{2\kappa^{2}} + p\,(\phi, \chi)
\right] \, , 
\label{action}
\ee
where $\kappa^2=8\pi G$, $\chi \equiv -\frac{1}{2}g^{\mu \nu}
\nabla_{\mu}\phi \nabla_{\nu} \phi$. The background homogeneous equation 
of motion is 
\begin{equation}
\Delta (\ddot{\phi }+3Hc_{s}^{2}\dot{\phi })+\frac{\partial
^{2}p}{\partial \phi \partial \chi }\dot{\phi }^{2}-
\frac{\partial p}{\partial \phi }=0\,,
\label{backgroundcomp}
\end{equation}
where
\begin{equation}
\Delta \equiv \frac{\partial p}{\partial \chi }+\dot{\phi 
}^{2}\frac{\partial
^{2}p}{\partial \chi ^{2}}\,, 
\label{delta} \quad \quad c_{s}^{2}\equiv 
\frac{\partial p}{\partial \rho } \bigg|_\phi=
\frac{\frac{\partial p}{\partial \chi }}{\Delta}\,,
\end{equation}
supplemented by the Hubble law
\be
3 H^2 = \kappa^2 \rho = \kappa^2 \left[ 2 \chi \frac{\partial 
p}{\partial \chi} - p \right] \;. 
\ee
We note that NEC is violated when $\partial p
/\partial \chi < 0$.
For simplicity we study scalar and tensor perturbations in the uniform 
curvature gauge around a flat Robertson-Walker (RW) line-element
\be
ds^2=-(1+2 \alpha)dt^2 - a \beta_{,i} dt dx^i+a^2 \left(\delta_{ij} 
+ h^{TT}_{ij}\right)dx^i dx^j
\,, 
\label{PER_UCG2}
\ee
The equation of motion for the scalar field fluctuation is
\begin{eqnarray}
& &\ddot{\delta \phi }+\left[ \frac{\dot{\Delta }}{\Delta }+3H\right]
\dot{\delta \phi }+\left[ -\frac{c_{s}^{2}}{a^{2}}\nabla^{2}-\frac{1}{\Delta
}\frac{\partial ^{2}p}{\partial \phi^{2}} \right. \nonumber \\
& & \left. + \frac{1}{\Delta a^{3}}\left(
a^{3}\frac{\partial ^{2}p}{\partial \phi \partial \chi }\dot{\phi }\right)
^{.}\right] \delta \phi \nonumber\\
& & = \left[\frac{\dot{\Delta }}{\Delta }\dot{\phi}+ 
2 \ddot{\phi} + 3H\dot{\phi} + 
3Hc_{s}^{2}\dot{\phi }\right] \alpha +\dot{\phi}\dot{\alpha}
-\frac{c_{s}^{2}}{2 a}\dot{\phi} {\nabla}^2 \beta \,.
\label{eqUCG}
\end{eqnarray}
By using the energy and momentum constraints from Einstein's equations 
\cite{BFM}, and going to Fourier space, we obtain
\begin{equation}
\ddot{\delta \phi }_{\bf k} + 
\left[ 3H+\frac{\dot{\Delta }}{\Delta }\right]
\dot{\delta\phi}_{\bf k} + 
\left[ c_{s}^{2}\frac{k^{2}}{a^{2}} - \frac{1}{a^3 \Delta x}
\left( a^3 \Delta \dot x \right)^.
\right] \delta \phi_{\bf k} = 0\,,
\label{gm2}
\end{equation}
where $x=\dot{\phi }/H$. This equation is one of the main results of our
paper. When $c_s^2 < 0$ fluctuations are unstable on small scales.
%and we shall not consider such a case. 
However, $c_s^2 > 0$ when $\Delta < 0$
and NEC is violated at the same time.
%this is the case we are 
%mainly interested in this Let.

By defining 
\begin{eqnarray}
v &=& a \delta \phi \sqrt{|\Delta|} \label{v} \\
z &=& a \frac{\dot \phi}{H} \sqrt{|\Delta|} = a \frac{|\rho + 
p|^{1/2}}{c_s H} \label{z}
\end{eqnarray}
we find that Eq. (\ref{gm2}) takes the form
\be
v''_{\bf k} + \left( c_{s}^{2}k^{2}-\frac{z''}{z}\right) v_{\bf k}=0\,.
\ee
The latter equation and the definitions in Eqs.~(\ref{v},\ref{z}) 
agree with those in Ref.~\cite{GM}, but are extended 
to the separate region where $\Delta < 0$ and $\rho + p < 0$.

For the configurations whose dynamics evolves across the boundary 
$\rho + p = 0$ \cite{vikman,hu,CD}, Eq.~(\ref{gm2}) multiplied by 
$\Delta$ can be used as a regular equation. The long wavelength solution 
for $\delta \phi_{\bf k}$ is:
\begin{eqnarray}
\delta \phi_{\bf k} &=& C(k) \frac{\dot \phi}{H} + D(k) \frac{\dot 
\phi}{H} 
\int d t \frac{H^2}{a^3 \dot{\phi}^2 \Delta} \nonumber \\
&=&  C(k) \frac{\dot \phi}{H} + D(k) \frac{\dot
\phi}{2 H M_{\rm pl}^2} \int d t \frac{c_s^2}{a^3 \epsilon_1}
\,,
\label{long}
\end{eqnarray}
where the second term is the decaying mode, and in the second line we have 
introduced $\epsilon_1 \equiv -\dot{H}/H^2$ and the reduced Planck mass 
$M_{\rm pl} = \kappa^{-1}$.

%comments on phantom divide crossing
The phantom crossing (if indeed occurs \cite{vikman})
affects $\phi$ fluctuations through $\Delta$ and not through $\partial p 
/\partial \chi$ (which is instead related to the parameter of state 
$w=p/\rho$). As it can be seen from Eq. (\ref{long}), if 
$\Delta \propto (t-t_*)$ in the
vicinity of the crossing $\dot H (t_*)=0$, the decaying mode diverges 
logarithmically at $t=t_*$. This fact has already been observed in a 
different gauge for theories with $c_s=1$ in \cite{CD}. Nevertheless, this 
divergence in field fluctuations may be softened (if not removed) in two 
cases: if $c_s^2$ 
also vanishes at the crossing (as it can be seen in the second line of Eq. 
(\ref{long})) or if $\Delta \propto (t-t_*)^{1/m}$, with $m>1$ being an 
odd integer. We also note that scalar metric perturbations
$\alpha$ and $\beta$: 
\bea
\frac{H}{a} \nabla^2 \beta &=&\kappa^2 \Delta \frac{\dot \phi^2}{H} \left( 
\frac{H}{\dot \phi}
\delta \phi \right)^. \nonumber \\
2H\partial _{i}\alpha &=&
\kappa^2 \frac{\partial p}{\partial \chi} \dot{\phi} \partial_{i}\delta
\phi \,.
\label{einsteinperturbed2}
\eea
remain finite across the transition $\Delta=0$ 
irrespective of possible singularities in $\delta \phi$.

In terms of the {\it horizon flow functions} $\epsilon_i$ (defined as 
$\epsilon_{n+1} = \dot{\epsilon_n}/\left(H \epsilon_n\right)$, for $i \ge 
2$) the potential can be written as:
\be
\frac{z''}{z} = F + G 
\label{potential}
\ee
where $F$ is the usual (canonical) expansion term  
\be
F = a^2 H^2 \left( 2-\epsilon_{1}+\frac{3}{2}\,\epsilon_{2}
-\frac{1}{2}\,\epsilon_{1}\,\epsilon_{2}+\frac{1}{4}\,\epsilon_{2}^{2}
+\frac{1}{2}\,\epsilon_{2}\,\epsilon_{3} \right)
\label{KG}
\ee 
and $G$ depends on the sound speed,   
\be
G = 2 \left( \frac{c'_s}{c_s} \right)^2 - \frac{c''_s}{c_s} 
- 2 a H \frac{c_s'}{c_s} - a H \frac{c'_s}{c_s} \epsilon_2 \;. 
\label{sound}
\ee
Eqs.~(\ref{potential},\ref{KG},\ref{sound}) agree with Ref.~\cite{WCW} for 
$\epsilon_1 > 0$, but their validity is extended to the region 
$\epsilon_1 < 0$, corresponding to NEC violating models. 

The amplitude $h^{TT}$ of gravitational waves satisfies
\be
\ddot h^{TT}_{\bf k} + 3 H \dot h^{TT}_{\bf k} + \frac{k^2}{a^2} 
h^{TT}_{\bf k} 
= 0 \,,
\label{massless}
\ee
with the usual prediction $n_T = - 2 \epsilon_1$. For super inflation 
gravitational waves have therefore a blue spectrum \cite{LM2,gg} with 
the slow-roll prediction $r = 8 c_s n_T$. 
\\

\emph{A Toy Model}. As the simplest example of super inflation 
we consider the case with an exponential potential
\be
\rho = 
%\frac{1}{3 M_{\rm pl}} \left[ 
\sigma_K \frac{\dot\phi^2}{2} + 
\sigma_V V_0 e^{-\lambda \frac{\phi}{M_{\rm pl}}} \,,
%\right] \,,
\ee
where $\sigma_K \,, \sigma_V = \pm 1$. 
For $\sigma_K=\sigma_V=1$ the stable phase-space trajectories 
for homogeneous cosmologies occur for $\lambda < \sqrt{6}$ \cite{hall}. 
Power-law inflation \cite{LM1} ($\lambda < \sqrt{2}$) is a local asymptotic 
attractor among inhomogeneous cosmologies \cite{MSS}. 

The case with $\sigma_K=\sigma_V=-1$ leads to solutions in Euclidean time 
and we do not consider them here.

The case with negative potential ($\sigma_K=-\sigma_V=1$) leads to 
the simplest single field realization of the Ekpyrotic scenario \cite{KOST}. 
It represents a contracting solution with ultra-stiff matter ($w>1$) 
which is stable for $\lambda > \sqrt{6}$ 
both at homogeneous \cite{HW,EST} and inhomogeneous level 
\cite{EST} and solves the horizon problem. 
Unfortunately, this single field Ekpyrotic 
model cannot lead to the observed nearly scale-invariant spectrum for 
cosmological fluctuations \cite{BF}.

The case $\sigma_K=-\sigma_V=-1$ leads to a stage of  
pole-like inflation:
\begin{eqnarray} \label{pole}
a(t) &\sim& (-t)^{\, p} \,, \quad t<0 \,, \quad \quad p<0 \;, \nonumber
\\
\phi (t) &=& \frac{2}{\lambda} M_{\rm pl} \log (-M_{\rm pl} t) \;, 
\\
V_0 &=& M^4_{\rm pl} \, p \, (3p -1) \,, \nonumber 
\end{eqnarray}
where $p=-2/\lambda^2$ (i.e. $p < 0$). Such a solution is 
characterized by a constant state parameter 
$w=-1+2/(3 p) < -1$ and $\dot H > 0$ (we remind that $H=p/t$, where
$t=0$ corresponds to the Big Rip singularity). \\

\emph{Classical Stability}. The stage of pole-like inflation in 
Eq.~(\ref{pole}) is stable for {\it any negative} $p$. 
It is stable within the RW backgrounds: 
it is easy to show that the phase points 
\be
\left( \frac{ \kappa \dot \phi}{\sqrt{6} H} \,, 
\frac{\kappa \sqrt{V}}{\sqrt{3} H} \right)
= \left( - \frac{\lambda}{\sqrt{6}} \,, \sqrt{1+\frac{\lambda^2}{6}} \right)
\ee 
are stable. This stability analysis can be extended to the case of 
inhomogeneous space-times following closely the analysis in 
Ref.~\cite{MSS}, by introducing
\begin{eqnarray}
\label{hexp}
ds^2 &=& -d t^2 + (-t)^{2p} h_{ij} \, dx^i \, dx^j \;, 
\nonumber \\
h_{ij} (t,{\bf x}) &=& a_{ij} + \sum_n b_{ij}^{(n)} \, (-t)^{n}  
\;, \\
\phi (t,{\bf x}) &=& \frac{2}{\lambda} M_{\rm pl} \log (-M_{\rm pl} t) - 
\sum_n 
\Phi^{(n)} (-t)^n \,, \nonumber
\end{eqnarray}
where $a_{ij} \,, b_{ij}^{(n)} \,, \Phi^{(n)}$ are arbitrary 
space-dependent functions, $n \in \left\{ k n_1 + l n_2 + m n_3 \right\}$, 
$n_1 \,, n_2 \,, n_3$ 
are non-negative integers, with at least a positive one, and $k \,, l \,, m$ 
are positive real numbers \cite{MSS}. Inserting the expansion of 
Eq.~(\ref{hexp}) in the Einstein equations, along 
the lines of Ref.~\cite{MSS}, we obtain four solutions, two of which are 
the residual gauge modes of the synchronous gauge. 

On the basis defined by $k \,, l \,, m$, we 
first consider $n=k$, obtaining $k=1-3p$ and 
\be
\Phi^{(k)} = - \frac{b^{(k)}}{2 \lambda} \,, \quad \tilde{b}^{(k)}_{ij} 
{\rm arbitrary} \;, 
\ee 
where $\tilde{b}^{(k)}_{ij}$ is the trace free part of the tensor 
$b^{(k)}_{ij}$. For $n=l$ we obtain $l=2(1-p)$ and (for $p \ne 
-1$)
\begin{eqnarray}
\Phi^{(l)} &=& \frac{P}{2 \lambda p (p+1)(2p-3)} \;, \nonumber \\
b^{(l)} &=& \frac{p-3}{2 (p^2-1)(2p-3)} P \;, \\
\tilde{b}^{(l)}_{ij} &=& \frac{\tilde{P}_{ij}}{p^2-1} \;, \nonumber  
\end{eqnarray}
where $P$ and $\tilde{P}_{ij}$ are the trace and traceless parts of the 
three-dimensional Ricci tensor associated to $a_{ij}$.
The case $n=m$ with the double solution $m=0,-1$ corresponds to a gauge 
mode which is not fixed by the synchronous gauge in Eq.~(\ref{hexp}) 
(we note that for power-law inflation the solution is also twofold with 
$m=0,1$).

Therefore we have demonstrated that phantom inflation with an 
exponential potential described by Eq.~(\ref{pole}) is a local 
attractor (towards the Big Rip singularity) among inhomogeneous 
space-times for {\em any} slope of the potential, at the classical level. 
This independence on the slope of the potential may be interesting for a 
theoretical motivation of exponential potentials \cite{HR}. 
As for ordinary power-law inflation \cite{MSS}, only the field becomes 
smooth while the metric retains the initial $a_{ij} (x_0)$, stretched to 
ultra-large cosmological scales. 
It would be of great interest if quantum effects \cite{CHT} 
lead to the avoidance of the singularity \cite{NO} and drive the Universe 
into a radiation dominated era. More in general, some physical mechanism 
- a second field, for instance - driving the Universe out of the 
inflationary epoch has to be invoked 
\cite{vikman,piaozhang}, as for power-law inflation. \\

{\em Predictions of the toy model}. It is interesting to study 
cosmological perturbations on such a stable background (see also 
\cite{piaozhang} for phantom inflation with a generic potential). 
For an exponential potential, both scalar and tensor fluctuations satisfy 
the equation for massless fields, Eq.~(\ref{massless}).

In the background given by Eq.~(\ref{pole}) 
the solution for $X$ ($X$ can be either the amplitude $h$ of 
gravitational waves or the Mukhanov variable $Q$) is
\be
X_k = A (-\eta)^\nu H_\nu^{(1)} (-k\eta)
\label{solution}
\ee
where $A$ is the normalization factor and the index $\nu$ 
of the Hankel function is given by
\be
\nu = \frac{1}{2} \frac{3p-1}{p-1} = \frac{3}{2} + \frac{1}{p-1} \,.
%\, , \quad A =
%e^{i(\nu+\frac{1}{2})\frac{\pi}{2}} \frac{\sqrt{\pi}}{2}
%(p-1)^\frac{p}{p-1} \, .
\label{nuindex}
\ee
The spectrum of fluctuations is blue tilted with respect to scale 
invariance, but the tilt is suppressed for large $|p|$. 

As it happens also in the Ekpyrotic model, gauge invariant scalar 
fluctuations satisfy an equation in which the long-wavelength instabilities 
of field fluctuations in rigid space-time are removed due to the opposite 
sign of the kinetic and potential terms: this means that a consistent 
gravitational embedding leads to a self-regulation of these flat space-time 
instabilities.  
 
Both Ekpyrotic scenario and the super inflation sketched here violate the 
condition $|p| \le \rho$ and lead to a singularity (in contrast with  
power-law inflation). 
It is therefore necessary to solve the graceful exit problem in both cases. 
However there is an important difference: 
the exit from super inflation needs to reverse sign to 
$\dot H$ to match with a radiation era, while in the Ekpyrotic case 
the graceful exit is characterized by a switch in the sign of $H$ (as for 
the Pre-Big Bang scenario \cite{pbb} in the conformal frame). 
This difference reflects in the 
perturbation sector: in super inflation for $p < -1$ (i. e. for 
$n_T < 1$) growing and decaying 
modes remain as such before and after the graceful exit, while in bouncing 
models growing and decaying modes may invert their role before and after the 
bounce. 

\emph{Gravitational waves with a blue spectrum}. The detection of 
the tensor contribution to CMB anisotropies is a major observational 
challenge. In Fig. 1 we show that a blue spectral index $n_T > 0$ 
{\em increases} the possibility of detection of gravitational waves, 
since the tensor power-spectrum (in temperature and polarization) 
is increased at intermediate multipoles $\ell$ 
(before the tensor contribution is cut-off by the decay of 
gravitational waves inside the Hubble radius), for a fixed  
tensor-to-scalar ratio on the largest scales. 
The tensor contribution to CMB anisotropies depend on $n_T$, as it can be 
seen from Fig. 1: for $r=0.38$ (a value compatible with 
the WMAP $2 \sigma$ bound $r=0.43$, with no running spectral 
indices \cite{spergel}) the tensor polarization signal in super inflation 
with $n_T=0.1$ is more than twice the standard inflation one with 
$n_T=-0.1$. For comparison we also show the difference in the tensor 
contribution between $n_T=0.9$ and $n_T=-0.1$ in Fig. 2: increasing 
$n_T$, the increase in polarization amplitude is larger than in the 
temperature. Moreover, we also note that the peak of the BB spectrum 
shifts to larger $\ell$ increasing $n_T$. Other  
inflationary models beyond Einstein theories (scalar-tensor theories or 
higher order gravity theories) may also display a blue spectrum for 
gravitational waves. \\

\begin{figure}
\begin{tabular}{c}
\includegraphics[scale=0.48]{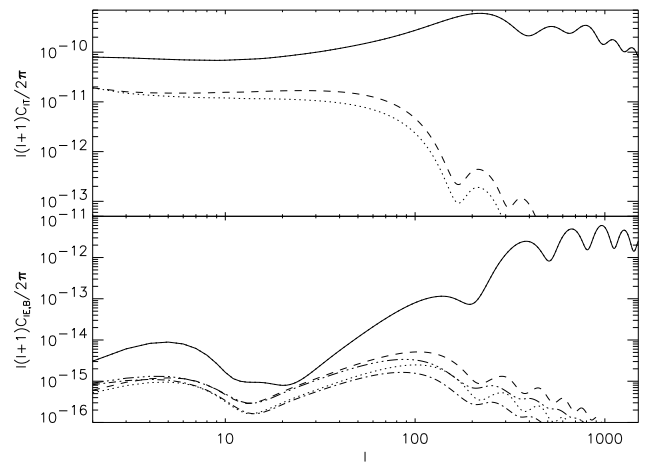}
\end{tabular}

\caption{Comparison of the CMB anisotropy power-spectrum for conventional 
inflation and super inflation. 
%Temperature in 
%the top panel and E and B-modes polarization in the right panel. 
For scalar perturbations a blue spectrum $n_S = 1.1$ is assumed 
(as predicted by hybrid models in conventional 
inflation) for both standard and super inflation. For tensors  
$n_T = -0.1$ is assumed for inflation and $n_T = 0.1$ for super inflation. 
The tensor-to-scalar ratio $r \sim 0.38$ has been assumed. 
For reionization, the subroutine {\sl recfast} has been used,  
with optical depth $\tau=0.1$. The other parameters used are 
$\Omega_{\rm CDM}=0.26$, $\Omega_{b}=0.04$, $\Omega_{\Lambda}=0.7$ and 
$H_0=72$ km s$^{-1}$ Mpc$^{-1}$.
In the top panel, besides the temperature scalar spectrum, the dashed 
(dotted) line corresponds to tensor modes in super (standard) inflation. 
In the bottom panel, besides the scalar E-mode (solid), the tensor E-mode 
is the dashed (dotted) line and the B-mode is the triple dot-dashed 
(dot-dashed) line for super (standard) inflation.}
\end{figure}

\begin{figure}
\begin{tabular}{c}
\includegraphics[scale=0.5]{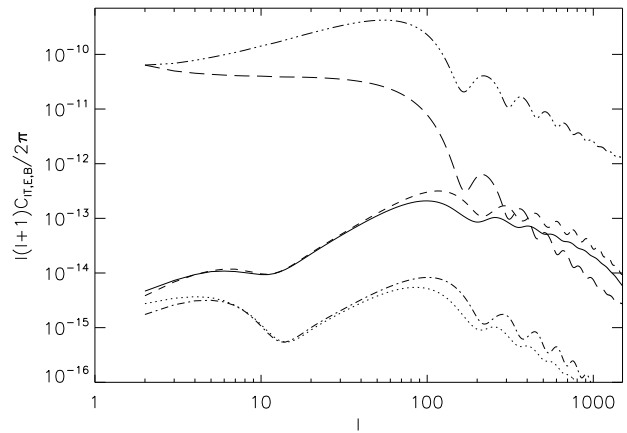}
\end{tabular}

\caption{Comparison of tensor contribution to CMB anisotropy for super 
inflation with $n_T=0.9$ and standard inflation $n_T=-0.1$ (with the same 
amplitude in temperature anisotropies on large scales). Besides the TT 
spectrum as a triple dot-dashed (long dashed) line, 
the EE spectrum is dashed (dot-dashed) and the BB one is solid 
(dotted) for super (standard) inflation. Note the different increase in 
the spectra and the generic shift at larger $\ell$ of the first peak 
when increasing $n_T$.} \end{figure}

%Gravitational waves with a blue spectrum may become observationally 
%important at intermediate 
%frequencies, in contrast with standard inflation. \\

\emph{Conclusions}. In this paper we have presented a consistent framework
to study super inflation and given exact solution for a toy-model
with an exponential potential. While these models deserve further 
investigation both in connection with quantum instabilities \cite{CHT}, 
non-perturbative effects and the exit from the accelerated stage, 
they provide the distinctive prediction that the 
tensor perturbation spectrum
is blue-tilted (independently on the number of scalar fields involved). 
This might also open new windows for the direct detection 
of the stochastic gravitational-wave background by interferometric 
antennas. 
Indeed, the ratio w.r.t. a scale-invariant tensor spectrum of the 
contribution to closure energy density 
in gravitational waves (per unit log frequency) scales like 
$(k/k_*)^{n_T} \sim (5 \times 10^{16})^{n_T}(k/{\rm Hz})^{n_T}$, if
the tensor spectrum is normalized to CMB observations at $k_* = 0.002$
Mpc$^{-1}$ \cite{spergel}. This may lead to a relevant increase of the 
signal for positive tilt, 
even without requiring large deviations from scale-invariance, 
unlike the Ekpyrotic scenario with $n_T=3$ or Pre-Big 
Bang with $n_T=4$. Given the completely general scalar-field theory
we have considered, such a blue spectrum for gravitational waves does 
not constrain the slope of the scalar perturbation spectrum, 
therefore allowing agreement with observational data.  \\

\emph{Acknowledgement}. We wish to thank A.~A.~Starobinsky for 
discussions. F. F. wishes to thank INFN for support.

\end{document}